\documentclass[aps,preprint,showpacs,groupedaddress]{revtex4-1}  
\usepackage[latin1]{inputenc}
\usepackage[T1]{fontenc}
\usepackage{graphicx}  
\usepackage{dcolumn}   
\usepackage{bm}        
\usepackage{amssymb}   
\usepackage{amsmath}   
\usepackage{psfrag}
\usepackage{epsfig}
\usepackage{appendix}
\def\1{\sigma \sigma^\prime}
\def\2{\bar{\sigma}\bar{\sigma^\prime}}
\def\U{U^{11}}
\def\V{V^{11}}
\def\Uone{U_1}
\def\Utwo{U_2}
\def\Vone{V_1}
\def\Vtwo{V_2}
\def\R{R}
\def\bD{\overline{D}}
\def\hD{\widehat{D}}
\def\S{{\cal R}}
\def\Sp{{\cal R}^\prime}

\def\cov{\rm Cov}
\def\Var{\rm Var}
\def\tr{\rm tr}
\def\E{\mathbb{E}}
\def\p{\pi}
\def\P{\Pi}
\begin{document}

\title{Generalized exchange cluster algorithm to compute efficiently covariances and susceptibilities in Monte Carlo}
\author{Roland Assaraf$^{a)}$, Hilaire Chevreau $^{a)}$}
\affiliation{
  $^{a}$ Laboratoire de Chimie Th\'eorique (UMR 7616), Sorbonne Universit\'e, CNRS, Paris, France
}
\date{\today}

\begin{abstract}
  We present a Monte Carlo method to compute efficiently susceptibilites  or covariances of two physical variables.
  The method  relies on a  generalization of the exchange cluster algorithm to any model of interacting particles with any $2$-body interactions.
  The principle is to select clusters of variables belonging to two independent replicas of the system. An improved estimator of the covariance of two physical variables (in one replica) is then proposed. This estimator has the zero-variance property in the limit where these variables are independent. 
 In practice the  scaling of the statistical fluctuations as a function of the number of degrees of freedom $N$ is reduced  from $O(N^2)$ to $O(N)$.
   This   lower  scaling is illustrated on a Lennard Jones model.
\end{abstract}
\maketitle
\section{Introduction}
At the center of the $N$-body problem in physics is the evaluation of physical properties which are usually logarithmic  derivatives of a large dimensional integral, a path integral  in quantum physics at zero or finite temperature,  a partition function in statistical physics.

The magnetization is a first derivative with respect to an external magnetic field, the energy is a first derivative with respect to a time (in quantum physics at zero-temperature), the free energy is a derivative with respect to  the inverse temperature $\beta$.
Response  properties like the magnetic susceptibility, the heat capacity or forces come out as second order derivatives.
Monte Carlo methods are techniques of choice to compute such properties but may still be computationally  demanding for a system with a larger number $N$ of particles.
It is very  known that the convergence of the sampling process  may be slow especially in the presence of long range correlations, at low temperature or close to a critical point (critical slowing down).
Statistical fluctuations may also be prohibitive for large systems, when computing  small energy differences or  second order derivatives (and beyond). 

Susceptibilities are second order derivatives of the partition function or first order derivatives of the energy, and can be expressed as  covariances  of usually extensive properties. The 
result is extensive but the variance  $\sigma^2$  is usually quadratic ($O(N^2)$) leading to numerical inefficiency for a large system even with short range correlations at high temperature.

Using a general zero-variance principle for computing properties \cite{assaraf:4682,assaraf_10.1063/1.1621615} can reduce fluctuations by orders of magnitude but the scaling as a function of $N$ of a first order response property (first order energy derivative or susceptibility) is an open problem \cite{assaraf_PhysRevE.89.033304}.
Some proposals exist like performing an explicit derivative of the stochastic dynamic with respect to parameters \cite{PhysRevE.90.063317}. However 
 this method is limited by the ability to use a non-chaotic stochastic process with a synchronization property which is not easy to build in the general case \cite{assaraf2011,Assaraf2017}. 
 
Here we propose to use the exchange cluster (or domain) algorithm to build an estimator of susceptibilities with size-extensive variances. 
The exchange-cluster algorithm consists in sampling a pair of independent replicas of the system and building domains across the two replicas, such that these domains can be exchanged between the replicas without modifying the stationary distribution.
This algorithm has already been employed on spin or generalized Pott models and exploited to compute efficiently a few specific thermodynamic properties \cite{Hasenbusch_Casimir_2013,Hasenbusch_Casimir_2015} and  two or three-point correlation functions  \cite{PhysRevE.97.012119,PhysRevE.96.032803}.
By efficient we mean here that the calculation of extensive properties has a linear complexity as a function of the system size $N$.
This also means that the variance of the estimator scales linearly with $N$. 
Here we propose two new developments to achieve this property: firstly an extension of the exchange cluster algorithm to any system, whether on the lattice or in the continuum space, secondly a general efficient estimator for susceptibilities. 
It turns out that these estimators are efficient in the regime when the number of domains scales linearly with the system size, which is always reached in the limit of weak coupling or high temperature.
The exchange cluster algorithm can be seen as an instance of the generalized geometric cluster algorithm \cite{Heringa1998GeometricCM,Liu2005-hu,Luijten2006} but applied to an extension of the system consisting in two independent replicas.
The geometric cluster algorithm generalizes the celebrated Swendsen Wang and Wolf algorithms originally developed for spin systems  \cite{PhysRevLett.57.2607,PhysRevLett.62.361,Chayes_stat_phys1998}.

This family of algorithms relies on a self inverse transformation which corresponds to a global symmetry of the Hamiltonian \cite{Heringa1998GeometricCM,Liu2005-hu,Luijten2006}  to build a global move that preserves the stationary distribution. Such self-inverse transformation can be a spin flip for spin systems, or a rotation of particles in the continuum space. In the exchange cluster algorithm, this transformation is the exchange of physical variables (spins, coordinates) between the replicas.

These methods are mainly employed to reduce the slowing down of the dynamic near criticality. This has the indirect effect to reduce the statistical fluctuations by lowering the autocorrelation time.
They have also employed in some instances to reduce directly statistical fluctuations associated to the sign problem in the family of Hubbard models \cite{PhysRevLett.83.3116} or to compute a few susceptibilities having the right symmetry  \cite{WOLFF1988501,HASENBUSCH1990581,WOLFF1990581,JANKE1998329}.
Here we propose to exploit the exchange-domain algorithm to draw a general efficient estimator for susceptibilities.
We point out that the exchange cluster algorithm is not the only cluster method using replicas, one of the most known and versatile  methods being the so-called replica exchange algorithm or the parallel tempering method \cite{geyer1991computing,Earl2005ParallelTT}.
With these techniques the full system can be exchanged between different independent replicas at different temperatures. Again this techniques  accelerate the dynamic but do not lower the scaling of the variances.





In a first section we introduce the algorithm which consists in building domains of pairs $(i,j)$ such that $i$ like $j$ is an index of a physical variable in each of the two replicas (for example $i$ may represent the position  of two particles, one in each replica). 
All pairs of variables in a given domain can be permuted between the two replicas, leaving the probability measure of the system invariant. 
Such property can be in principle used to improve the ergodicity in the simulation.
The focus here is that thanks to this property, we can build improved estimators of susceptibilities, i.e. a covariance written as an effective energy interaction between the domains. The scaling of the variance is then reduced from $O(N^2)$ to $O(N^2 / N_D)$ where $N_D$ is the number of domains.

This  generalized cluster-exchange algorithm extends a previous realization  on the lattice  \cite{Hasenbusch_Casimir_2015}.  It is also more efficient than other previous versions  of algorithms exchanging clusters between two independent replicas \cite{Houdayer_2001} in the sense that it systematically increases the number of clusters (or domains) that can be exchanged.

In the second section we present numerical tests on the well- known Lennard Jones model.

\section{Extension of the exchange cluster algorithm to a general statistical model}

\subsection{Statistical models and complexity of computing the response functions}
We consider  a general probability density written as follows
\begin{equation}
  \p(\R) = \prod_{I \subset \Omega} w_I(\R_I)
  \label{genmod}
\end{equation}
where $\Omega$ is a set of indices $i$ and $\R = (\R_i)_{i \in \Omega}$ is a list of variables that characterizes a configuration of the physical system.
$\R_I =  (\R_i)_{i \in I }$  is a subset of variables ($I \subset \Omega$) and  $w_I(\R_I)$ is a positive function of $\R_I$.
For instance in the Ising model $i$ is an indice of a point on the lattice and 
$\R_i = S_i = \pm 1$ is the spin on that point, $w_{\{i,j\}} (S_i, S_j) = e^{-\beta S_i S_j}$ if $i,j$ are neighbors and $w_I=1$ for any other subset of variables.
Conversely for a model of interacting particles in real space
 $i$ is a label for a particle and $\R_i$  represent its three spatial coordinates. For a pair-interacting model like a Coulomb potential, the subsets are  pairs of indices and $w_{ij} (\R_i, \R_j) =e^{-\beta \frac{1}{r_{ij}}}$   where $r_{ij}$ is the distance between particles $i$ and particle $j$. 
Kinetic energy terms can also be represented with this equation, velocities appearing as additional  variables which can be coupled or not via $w_I$.
This expression is then quite general in statistical physics, it can also be  applied for quantum bosonic systems at finite temperature or at zero temperature using and imaginary time dynamics. Of course we don't consider here  quantum fermionic problem  for which Eq.~(\ref{genmod}) holds but  $w_I$  is not positive.

Let's begin with a general two-body interacting system
\begin{equation}
  \p(\R) = \prod_{i,j} w_{ij}  (\R_i,R_j)
  \label{pprodw}
\end{equation}
and a response function which thanks to the fluctuation dissipation theorem is a covariance of two properties  
\begin{equation}
\chi = {\rm cov}(U,V) \label{chidef} 
\end{equation}
with 
\begin{eqnarray}
U &= & \frac{1}{2}\sum_{i, j} u_{ij}  \nonumber \\
V &=& \frac{1}{2}\sum_{i, j} v_{ij}
\label{covuv}
\end{eqnarray}
where  $u_{ij}= f_{ij}(\R_i,\R_j)$, $v_{ij}= g_{ij}(\R_i,\R_j)$ are two-body symmetric interactions if $i \ne j$,   $u_{ii}/2$ and $v_{ii}/2$ are  one-body interactions.
Let's evaluate the numerical complexity to compute $\E(U)$, $\E(V)$ and $\chi$. We remind that the straightforward estimator of the expectation value of X, $\E(X)$ on a sample of size $M$ is the statistical average    
\begin{equation}
\overline{X} \equiv \frac{1}{M} \sum_i X_i 
\label{stataverage}
\end{equation}    
The numerical error is proportional to 
$\sigma(X)/\sqrt{M}$  where $\sigma$ is the standard deviation i.e. $\sigma(X) = \sqrt {\Var(X)} = \sqrt{\cov(X,X)}$.  

 Despite the fact that the number of terms in $U$ and $V$ is quadratic in $N$, they involve interactions between particles  which vanish at long distance.  Consequently if the system is sufficiently large, $U$ and $V$ can be decomposed as sums of terms corresponding to $N_D$ weakly interacting fragments. This should reduce the quadratic complexity.
 This motivates to assume the independent fragments model
 \begin{eqnarray}
 U &=& \sum_{I=1}^{N_D} U_I \nonumber \\
 V & = &  \sum_{I=1}^{N_D} V_I \label{UVindependent}
 \label{defUV}
 \end{eqnarray}
 where $U_I$ is independent of $U_J$ and $V_J$ for any $I \ne J$.
The consequence is that the expectation values of $U$ and $V$ are size extensive, i.e. they grow linearly with the system size.
The covariance is also size-extensive since $\cov (U ,V) = \sum_{I,J} \cov(U_I,V_J) = \sum_I \cov (U_I,V_I)$. The last simplification comes from the independence of $U_I$ and $V_J$ if $I \neq J$.
As a result the variances of $U$ and $V$ which are respectively $\sigma^2(U)$ and $\sigma^2(V)$ are size-extensive like $U$ and $V$.  
Using $U$ and $V$ as estimators of their respective expectation values $\E(U)$ and $E(V)$ has a numerical efficiency independent of the system size: the ratio signal / noise  proportional to  ${\E(U) \sqrt{M}}/{\sigma(U)}$  is not increasing with $N$.

Conversely for a response function the variance of the usual estimator $UV$ (assuming without reducing generality that $U$ and $V$ are centered) 
is $O(N^2)$ leading to a ratio signal / noise in $O(1/N)$. 
To prove it, we evaluate the variance ${\rm Var}(UV) = \E(U^2V^2) -\E(UV)^2$. When expanding $U$ and $V$ in their $N_D$ terms leads to $N_D^2$ and using the independence of the fragments  we find 
$\Var(UV) = \Var(U)\Var(V)+2\cov(U,V)^2-2 \sum_I (U_I V_I)^2-\E(UV)^2$ which involves products of size extensive terms, this ends the proof that $\Var(UV)=O(N^2)$.
Indeed, contrary to $\Var(U)$, $\Var(UV)$ is not size extensive because  $(UV)^2$ involves $O(N^2)$ positive terms  of the form $(U_I V_J)^2$ between different fragments $I$ and $J$. 
In conclusion the standard estimator .Eq~(\ref{chidef}) of a covariance is  inefficient when  
 $N$ is large. This is of course a bottleneck when exploring the thermodynamic limit.
The idea is then to build an estimator of the response function $\chi$ with the same form as $U$ or $V$ in ~Eq.~(\ref{covuv}) to insure the size-extensivity of its variance 



\subsection{Generalized exchange cluster algorithm} \label{genexch}
In the following we will define fragments which allow us to write such an estimator of $\chi$.
Here we propose to perform a stochastic simulations on two independent replicas of the system, i.e. we  sample the density
\begin{equation}
 \P(\S)= \Pi(\R^1, \R^2) = \pi(\R^1)\pi(\R^2)
  \label{pairdens}
\end{equation}
where $\S=(\R_1,\R_2)$ represents a pair of configurations of the independent replicas.
We point out that sampling the density is equivalent to perform a stochastic process collecting  configurations 
$\S_k=(\R^1_k,\R^2_k)$ verifying the ergodic theorem, i.e. for any continuous function $f(\R)$  
\begin{equation}
\lim_{M \to \infty} \frac{1}{M} \sum_{k=1}^M f(\S_k) = \E(f(\S)) \equiv \int f(\S) \P(\S) d\S
\nonumber
\end{equation}
Note that the expectation value $\E(f)=\E(f(\S))$ is obtained with $\S$ distributed according to the density $\Pi$.
Formally we can also write with Dirac notations
\begin{equation}
\lim_{M \to \infty} \frac{1}{M} \sum_{k=1}^M \langle \S_k \mid = \E( \langle \S \mid) \equiv \int   \Pi(\S) \langle \S \mid d\S
\nonumber
\end{equation}
where  $\mid \S \rangle$ represents the Dirac distribution centered on $\S$, i.e. $\langle \S^\prime \mid \S \rangle= \delta(\S^\prime -\S)$. We now define domains in such a way that exchanging the particles between the two replicas leaves the density $\P$ invariant.
We remind that $\Omega$ is the set of indices of the variables defining a configuration of the system, for example $i$ is the index of a particle in the Lennard-Jones model, but also the index of a pair of two particles belonging to the two replicas. Of course this supposes that a numbering of particles and a correspondence between the particles in the two replicas have been chosen.
A partition of domains $D_m \subset \Omega$ is selected as follows. Each pair of indices $\{i,j\}$ is not linked according to the Metropolis 
probability
\begin{equation}
  q_{ij} = \min \left(\frac{w_{ij}^{12}w_{ij}^{21}}{w_{ij}^{11}w_{ij}^{22}} ,1\right)
\label{prob_link}
\end{equation}
or linked with the probability $1-q_{ij}$. Note that in the Lenard Jones example the pair  $\{i,j\}$ represents two pairs of particles belonging to the two replicas.
In this expression (\ref{prob_link}) we introduced the following simplified notation: 
for any function $f_{ij}(R_i,R_j)$ and any pair $(\sigma,\sigma^\prime) \in \{1,2\}^2$
\begin{equation}
  f_{ij}^{\sigma\sigma^\prime} \equiv f_{ij} (\R_i^\sigma,\R_j^{\sigma^\prime}) \
\end{equation}
where the last equality comes from the symmetry of the interaction. The denominator in Eq.~(\ref{prob_link}) represents the contribution of the variables (or the particles) of indices i and j in the two replicas to the full interaction. The numerator represents the contribution coming from the same variables when only one variable for example with subscript $i$ has been exchanged between the system 1 and 2. 
Hence a large Metropolis weight increases the probability to swap the particles corresponding to one index but not the other ($i$  or $j$) in the two replicas.  The indices $i$ and $j$ are then not linked in this sense, the corresponding particles can be permuted independently. This occurs for all the pairs  at high temperature $T$ since $\lim_{T \to \infty} q_{ij}=1$. 
Conversely a small Metropolis weight increases the probability of the indices to be linked, and prevents the exchange of the corresponding particles separately but both of them. This occurs for all the pairs  at low temperature since $\lim_{T \to 0} q_{ij}=0$ (indeed is the system is frozen in a minimal energy state almost all the moves increase the energy and are forbidden).         
We are now building domains of indices such that the corresponding particles can be exchanged between the two replicas without modifying the measure.
The algorithm draws a set $B$ of all linked pairs of indices, these pairs correspond to the edges of a graph. From this set, maximally linked domains $D$ are built, in which all the indices of $D$ are linked directly or indirectly (maximal graph components). 

For any domain $D$ belonging to this list of domains determined by $B$, we introduce the exchange-domain  operator $\hD$ which applied on  the configuration of the full system $\S \equiv (\R^1,\R^2)$ swaps the coordinates of the particles in the domain $D$ between the two replicas. 

Formally using Dirac notations the application of $\hD$ to $\mid \S \rangle=\mid \R^1,\R^2\rangle$ gives  $\hD \mid \S \rangle = \mid \Sp \rangle = \mid \R^{\prime 1},\R^{\prime 2} \rangle $
 such that
 \begin{eqnarray}
 R^{ \prime 1}_D & =&  R^2_D \nonumber \\
 R^{ \prime 2}_D &=& R^{1}_D \nonumber \\
R^{ \prime 1}_{\bD} &= &R^1_{\bD} \nonumber \\
R^{ \prime 2}_{\bD}&=& R^{2}_{\bD} \nonumber
\end{eqnarray}
where $\bD$ is the complementary domain of $D$ in $\Omega$ and $\R^1_D$ (resp $\R^2_D$) is the restriction of $\R^1$ (resp $\R^2$) to the domain $D$, i.e. $\R^1_D$ (resp $\R^2_D$) are the coordinates of the particles with indices belonging to the domain $D$ in the first replica (resp the second replica).
 
The fact that the transformation $\hD$ leaves the distribution $\P$ invariant, is a consequence of the detailed balance property 
\begin{equation}
   \P(\Sp)  P(\S \mid \Sp) =  \P(\S)  P(\Sp \mid \S)
   \label{detbal}
\end{equation}
Let's prove this property.
First we know from Eqs.~(\ref{pprodw},\ref{pairdens}) that 
\begin{equation}
\Pi(\S)=\prod_{ij} w_{ij}^{11} w_{ij}^{22} 
\nonumber
\end{equation}
Remarking that $\Pi(\Sp)$ differs from $\Pi(\S)$ only by cross domain terms $(i,j) \in  D\times \bar{D}$ we can write   
\begin{equation}
\P(\Sp)   = \P(\S) \prod_{{(i,j)  \in  D\times \bar{D}}} \frac {w_{ij}^{12}w_{ij}^{21}}{w_{ij}^{11} w_{ij}^{22}} \label{pd}
\end{equation}
   
The  probability to draw $B$ from any configuration $\S$ is 
$$P(B|\S) = \prod_{\{kl\} \subset B} (1-q_{kl}) \prod_{\{ij\} \subset \bar{B}} q_{ij}$$
It is also differing from  $P(B|\Sp)$  by cross-domain terms: $q^\prime_{ij} \ne q_{ij}$ only if $(i,j) \in  D\times \bar{D}$.
Consequently 
\begin{eqnarray}
  P(B|\Sp) & =& P(B|\S )
  \prod_{(i,j) \subset  D\times \bar{D}} \frac{q_{ij}^\prime}{q_{ij}} \label{pbd}
\end{eqnarray}
Note that two (directly) linked indices $(k,l)$ ($\{kl\} \subset B$)  are necessarily in the same domain and do not change from $\S$ and $\Sp$. That's why the terms of the form  $1-q_{kl}$ with $\{kl\} \subset B$ do not appear in Eq.~(\ref{pd}).
Using Eq.~(\ref{prob_link}) the  terms in the products Eq.~(\ref{pd},\ref{pbd}) appear to be the inverses of each other.
Combining Eq.~(\ref{pd},\ref{pbd}) by a simple product side by side  we obtain the expression
\begin{equation}
  \P(\Sp)  P(B|\Sp) =  \P(\S) P(B|\S)
  \label{detbalB}
\end{equation}
If we can replace $B$ by $\Sp$ in the r.h.s of Eq.~(\ref{detbalB}) and by $\S$ in the l.h.s we recover the detailed balance property Eq.~(\ref{detbal}). This replacement can be done because the knowledge of B given $\S$ leads to $\Sp$ and reciprocally the knowledge of B given $\Sp$ leads to $\S$.
This idea is expressed in the following last step of the proof. 
 Evaluating the r.h.s. of Eq.~(\ref{detbal}) and using the law of total probability we find 
 \begin{eqnarray}
  \P(\Sp)  P(\S |\Sp)
                      & = &  \sum_B \P(\Sp)  P(\S | B, \Sp) P(B | \Sp)  \nonumber \\ 
  &=& 
                   \sum_B \P(\S) P (\S |B, \Sp) P (B |\S) \label{ptrans_exp}  
\end{eqnarray}
where the last identity is obtained using Eq.~(\ref{detbalB}) .
We make the symmetry hypothesis, that given  $B$,  $\S$ and $\Sp$ are related with the same function $f$, i.e.  
\begin{eqnarray}
\S = f(B, \Sp)  \ \Leftrightarrow   \Sp = f(B, \S)
\label{sym}
\end{eqnarray}

Consequently the conditional probability $P(\S | B, \Sp)$ is a Dirac function and 
\begin{eqnarray}
P(\S | B, \Sp) &=&\delta (\S - f(B, \Sp)) \nonumber \\
& =& \delta (\Sp - f(B, \S)) \nonumber \\
& =& P(\Sp | B, \S) \nonumber
\end{eqnarray}
Using this symmetric property to rewrite Eq.~(\ref{ptrans_exp}), followed by the law of total probability one recovers the detailed balance  property
\begin{equation}
   \P(\Sp)  P(\S \mid \Sp) =  \P(\S)  P(\Sp \mid \S)
\end{equation}
which ends the proof that the stochastic process leaves $\Pi$ invariant.

In practice such property means that we can swap all the variables (or the particles) inside any domain without modifying the stationary distribution (\ref{pairdens}). 
Such process can be added as a step to sample this distribution and improve the ergodicity of the algorithm like the Swendsen Wang or Wolf algorithms \cite{PhysRevLett.57.2607,PhysRevLett.62.361}. We  focus here on using this exchange domain  $\hD$-invariant property to build improved estimators of response functions like in  Eq.(\ref{covuv}).
We first enlight a nice property of the domains. In the separability limit where the physical system is a combination of two independent subsystems indexed by $A$ and $B$, the set of domains separate $A$ and $B$ i.e. $A$ and $B$ are both (disjoint) unions of domains. 
Indeed if there is no interaction between any pair of particles $(i,j) \in A \times B$, $w_{ij}^{12}=w_{ij}^{11}=w_{ij}^{22}=1$ and $q_{ij}=1$. This means that any pair of indices belonging to $A$ and $B$ are unlinked with a probability one. 
Consequently in a such separable system there are at least two domains.
In general two fragments $A$ and $B$ have the probability $\prod_{(i,j) \in A \times B} q_{ij}$ to be domain-separated. The weaker the interaction between two subsystems is, the larger is their probability to be domain-separated. 
In the limit where the system can be approximated by a extensive number of independent fragments, the number of domains should be extensive. 
This property of separability of domains is at the core of building an estimator of the covariance which is zero (with a variance zero) in the limit where the two random variables are independent.

One can check that the algorithm formulated by Hasenbusch on the Blume Capel model \cite{PhysRevE.93.032140} can be viewed as a specific realization of this general method.
The later are more efficient than the exchange cluster method previously developed for spin systems 
\cite{Houdayer_2001} for which it's easy to check that the criteria to unlink a pair is much stronger: 
the unlinked pairs in this previous reference are a subset of the unlinked pairs obtained in the present algorithm, and correspondingly the domains of this previous reference are an union of the domains obtained with the present work.

Note that this statement of efficiency looks like to be universal regarding the general geometric cluster framework \cite{Liu2005-hu}.
For example if the self inverse transformation is reverting two spins with the same index in two replicas, it's easy to check that the reference \cite{PhysRevLett.57.2607} provides a stronger criteria to unlink a pair of indexes than a direct application of the geometric cluster algorithm (with the same transformation).
But of course here we focus only on one transformation, the exchange of variables between replicas which will provide a general variance reduced estimator for response functions.


\subsection{Improved estimators for the response functions}
The following expression is an unbiased estimator of $\chi=\cov(U,V)$ where  $(U,V)$ are properties on the system Eq.~(\ref{defUV}).
\begin{equation}
\frac{1}{2}(\Uone-\Utwo)(\Vone-\Vtwo) 
\label{chiestim1}
\end{equation}
where $(\Uone,\Vone)$ are the values of these properties on the first replica and $(\Utwo,\Vtwo)$  the values on the second replica.
This can be easily checked expanding the product and computing the expectation value 
\begin{equation}
\frac{1}{2}(\E(\Uone \Vone) + \E(\Utwo \Vtwo)) =\E(\chi)
\end{equation}
where we used the independence of the two replicas that cancels the cross terms like $\E(U^1 V^2) = \E(U^1) \E(V^2)=0$ assuming all the variables are centered without losing in generality.
The estimator (\ref{chiestim1}) involves a product of two variations of potentials between the two replicas.

We will decompose these variations $U^1-U^2$ and $V^1-V^2$ as a sum of effective interactions between two domains $D_m$ and $D_n$  for any $(m,n)$ and use this decomposition to express $\chi$.
The effective interactions between different domains $D_m$ and $D_n$ are  
\begin{eqnarray}
  U_{mn}^{\1} &\equiv& \sum_{(i,j) \in D_m\times D_n} u^{\2}_{ij} -u^{\1}_{ij}  \nonumber \\
  V_{mn}^{\1} &\equiv& \sum_{(i,j) \in D_m\times D_n}v^{\2}_{ij} - v^{\1}_{ij}
  \label{antUV}
\end{eqnarray}  
where $\bar{2}\equiv 1$, $\bar{1} \equiv 2$.
With $m \ne n$, $U^{11}_{mn}=-U^{22}_{mn}$ represents the variation of the inter-domain $(m,n)$  interaction between the two replicas. 
If $m=n$, we consider the same expression ~(\ref{antUV}) but because of the double counting over particles, $U^{11}_{nn}/2$  represents a one-domain contribution: the one-domain ($n$) energy difference between the two replicas or   
equivalently the energy variation in the domain $m$ in the system 1 after the exchange of particles in this domain between the two replicas.
Note that these effective interactions are anti-symmetric $U_{mn}^{\1}= -U_{mn}^{\2}$.

With this definition (\ref{antUV}) the variation 
$U^1-U^2$ becomes a sum of inter-domain interactions: 
\begin{eqnarray}
\Uone-\Utwo & =& \frac{1}{2} \sum_{i,j} u^{11}_{ij}-u^{22}_{ij} \nonumber \\
        & =& \frac{1}{2} \sum_{mn} \sum_{(i,j) \in D_m \times D_n} u^{11}_{ij}-u^{22}_{ij} \nonumber \\
        & =& \frac{1}{2} \sum_{mn}  U^{11}_{mn} 
\end{eqnarray}
and the same expression holds for $\Vone-\Vtwo =  \frac{1}{2} \sum_{mn}  V^{11}_{mn}$.

Expanding the expression (\ref{chiestim1}) we find
\begin{equation}
\frac{1}{2}(\Uone-\Utwo)(\Vone-\Vtwo) = 
\frac{1}{8} \sum_{m, n, p, q} U_{mn}^{11} V_{pq}^{11} .
\label{domcom}
\end{equation}
This expression recovers in the language of domains that the estimator of $\chi$ is a sum of four-body interactions and its variance is $O(N^2)$, considering that a finite effective number of domains $n$ are interacting through $U$ and $V$ with any domain $m$.

The reduction of the scaling of the variance will come from the removal 
of all the unlinked terms i.e. the terms such that $\{m,n\} \bigcap \{p,q\}=\emptyset$, since as we prove later they have a zero-expectation value.
This leads to this new estimator of $\E(\chi)$
\begin{eqnarray}
\tilde{\chi} = \frac{1}{2}
\sum_{m}   {\cal U}_{m}^{11}{\cal V}_{m}^{11} -\frac{1}{2} \sum_{m<n} {U}^{11}_{mn} {V}^{11}_{mn}
\label{tchi}
\end{eqnarray}
where we introduced the interactions of the domain $D_m$ with all the domains 
\begin{eqnarray}
  {\cal U}_m^{\sigma \sigma^\prime} &\equiv&   \frac{U^{\1}_{mm}}{2} + \sum_{n \ne m}  U_{mn}^{\sigma \sigma^\prime} \nonumber \\
  {\cal V}_m^{\sigma \sigma^\prime} &\equiv& \frac{V^{\1}_{mm}}{2} + \sum_{n\ne m}  V_{mn}^{\sigma \sigma^\prime} 
    \label{defcal}
\end{eqnarray}

This new estimator Eq.~(\ref{tchi}) reformulates the response function  (a four body term) as a two body interaction (between domains), like $U$ and $V$  Eq.~(\ref{covuv}). As the expected result the variance of $\tilde{\chi}$ has the same scaling as the variances of $U$ and $V$ asymptotically $O(N)$, down from the original scaling $O(N^2)$.
This statement assumes that the average size of these domains does not depend on the size of the system beyond the correlation length, in accordance with the last paragraph of the former subsection.
We point out that this variance reduction comes with no additional computational complexity since computing $\tilde{\chi}$ and $\chi$ has the same bottleneck i.e. loops on the pairs $\{i,j\}$.
The estimator ~(\ref{tchi}) can be compared to the standard estimator computed on only one replica
\begin{equation}
  \chi = 
  U V -\E(U) \E(V)
  \label{chibare}
\end{equation}
Such evaluation of the second term of Expr.~(\ref{chibare}) is done using the product 
$\overline{U}\overline{V}$  involving a bias   $\E(\overline{U}\overline{V})-\E(U)\E(V)$ which is proportional to $1/M$. 
Hence besides the expected variance reduction for a large number of domains, another benefit of the exchange domain estimator ~(\ref{tchi}) is to remove this  bias.
However in the one-domain limit the variance of the exchange domain estimator is larger than the standard one. In this limit  $\Var(\tilde{X}) = \Var(X) + \Var(U) \Var(V)$ (see the appendix).

We now prove that $\tilde{\chi}$ is indeed an unbiased estimator of $\chi$. 
First $\hat{D}$ is self-adjoint  because for any function $f(\S)=\langle \S | f\rangle =  f(R^1,R^2)$ we can write  $\langle  \hD(\S) \mid f \rangle = \langle \S | \hD(f) \rangle=f(R^2,R^1)$.
Second we already know that $\Sp=\hD(\S)$ has the  density   $\Pi$ like $\S$. This implies  that for any function $f(\S)$,    $f(\hD(\S))$, i.e. the random variable  $\hD(f)$) has the same expectation value as $f$. 
Therefore the domains $D_m$ provide a basis for a large commutative algebra of operators  which acting on any observable  does not modify the expectation value.
We can choose for each term $U^{11}_{mn} V_{pq}^{11}$ in the expansion (\ref{domcom}) a specific operator which depends on the domains indexed by  $(m,n,p,q)$. This choice is equivalent to select for each term of the expansion~(\ref{domcom}) an operator which depends on the particles indices. The symmetry requirement (\ref{sym}) holds because the indices and the domains do not change after the exchange procedure.

We then choose to apply an operator (different from identity) only on the terms  of the form  $U^{11}_{mn} V_{pq}^{11}$ where  $\{m,n\} \cap  \{p,q\}=\emptyset$.
We apply on such a term the  operator  $\hat{L}_{mn} \equiv 1/2 (\hat{I} + \hat{D}_{mn})$ where $\hat{I}$ is the identity operator and $\hat{D}_{mn}$ is the exchange operator for the union of the domains $(D_m,D_n)$ (algebraically if $m=n$,  $\hat{D}_{mn}=\hat{D}_m$ and if $m \ne n$,  $\hat{D}_{mn}=\hat{D}_m \hat{D}_n$).
Since $\hat{D}_{mn}$ preserves the expectation value so is $\hat{L}_{mn}$.
Furthermore  since $\{m,n\} \cap  \{p,q\}=\emptyset$ we have $\hat{L}_{mn} (\U_{mn} \V_{pq})=0$. This is because $\hat{D}_{mn}( \U_{mn} \V_{pq} =\U_{mn} \V_{pq}=-\U_{mn} \V_{pq}$. 
In conclusion the unlinked terms have a zero expectation value so we can keep only the linked terms in Eq.~(\ref{domcom}) leading to the improved estimator 
\begin{equation}
  \tilde{\chi} = \frac{1}{8} \sum_{ \{m,n\}  \cap \{p,q\} \neq \emptyset} U^{11}_{mn} V_{pq}^{11}
  \label{op}
\end{equation}
To recover the  expression ~(\ref{tchi}) of the estimator we decompose Eq.~(\ref{op}) according to the number of different indices 
\begin{eqnarray}
  \tilde{\chi}
  & = &   \frac{1}{2}  \sum_{m \ne n, m \ne p, n \ne p}  \U_{mn} \V_{pm}         \nonumber  \\
  & &+\frac{1}{4} \sum_{m \ne  n}  \U_{mn} \V_{mn}+\U_{mm} \V_{mn} +  \V_{mm} \U_{mn}  \nonumber
  \\
   &  & +  \frac{1}{8} \sum_m\U_{mm} \V_{mm}  \label{decomp}  
\end{eqnarray}
where we used the symmetry $\U_{mn}=\U_{nm}$.  To prove  that this expression corresponds to Eq.~(\ref{tchi}) we use a matrix notation.
We decompose the matrices $\U$ and $\V$ in their diagonal ($U_d$ and $V_d$) and non diagonal ($U_c$ and $V_c$) parts. We have $\U=U_d+U_c$ and $\V = V_d + V_c$. 
Introducing the vector $|1 \rangle$\ which associates the number one to each domain $m$ $\langle m |1 \rangle=1$ we can rewrite the three terms of Eq.~(\ref{decomp}) as  
\begin{eqnarray}
\tilde{\chi}&=&\frac{1}{2} \langle 1 |U_c V_c |1 \rangle 
 -
\frac{1}{2}\tr(U_c V_c)`\\
& &+\frac{1}{4}\tr(U_c V_c)+\frac{1}{4}\langle 1 |\U_d V_c+U_c V_d| 1 \rangle  \\
& &+ \frac{1}{8}\langle 1 |\mid U_d V_d |\mid 1 \rangle
\end{eqnarray}

This expression can be rewritten after factorization  
\begin{eqnarray}
\tilde{\chi}&=&\frac{1}{2} \langle 1 \mid \left(U_c+\frac{U_d}{2}\right)\left(V_c + \frac{V_d}{2}\right) \mid 1 \rangle 
 -
\frac{1}{4}\tr(U_c V_c)
\end{eqnarray}
which is the matrix form of the expression (\ref{tchi}). 


\section{Application on the Lennard Jones model}
We now illustrate numerically the method on a Lennard Jones model. The Lennard Jones potential is 
\begin{eqnarray}
u_{ij} = 4 \epsilon [ (\frac{\sigma}{r_{ij}})^{12} - 
(\frac{\sigma}{r_{ij}})^{6} ]  
\end{eqnarray}
where $r_{ij}$ is the distance between particle $i$ and $j$, 
the total energy is $U= \frac{1}{2}\sum_{i, j} u_{ij} = \sum_{i< j} u_{ij}$.
The heat capacity $C_v$ is the variance of $U$ up to  the factor $k/T^2$  \cite{bennett_cv}  
\begin{eqnarray}
  C_v   \equiv  \frac{k}{T^2} ( <U^2 > - < U >^2) 
\end{eqnarray}
Of course the variance of $U$ is the covariance of $U$ with  $V=U$.
We will compute it either with  the usual estimator (\ref{chibare}) $T^2 Cv/k= \E({\chi})$    
or the exchange-domain estimator (\ref{tchi})  $T^2 Cv\_{\rm dom}/k = \E(\tilde{\chi})$ 
.  
We first test numerically some properties of the domains involved in the computation of $\tilde{\chi}$. 
We remind that the exchange-domain procedure is determined by the probability to unlink a pair of indices $(i,j)$ of variables (here particles positions) belonging to the two replicas.
Implicitly a correspondence is done between the particles of the two systems which share the same index. Note that particle indexes are not physical properties of the system,  they are just arbitrary names given to the particles which can depend on the particles positions. As a consequence any of these arbitrary choices does not change the detailed balance property and does not introduce any bias to the exchange-domain algorithm.  
However such a choice has an impact on the outcome of this algorithm since two particles with the same index are always attached to the same domain however far they can be. Consequently, the algorithm efficiency should depend on the correspondence between the  orderings of the particles in the two replicas, i.e. the list of pairs of particles sharing the same index. Such list or such pairing can be selected using any permutation $p$ of the particles in one replica where $p$ may explicitly depend on the configuration of the full system.  
We tested two criteria of pairing, the first criteria is a random order of the particles in both systems, we keep the same order at each step of the simulation allowing in principle fully spatially delocalized domains.
The second criteria  minimizes the  distance between the two systems $\sum_{i} |{\bf r}^1_i-{\bf r}^2_{i}|$. The permutation  $p$  is the outcome of the so-called assignment problem which is solved by the Hungarian algorithm \cite{kuhn_hungarian,munkres_hungarian,miller}.
With such criteria any particle $i \in R_1$ is assigned to a particle
$i \in R_2$ which should be spatially close. By contrast the second criteria favors spatially localized domains. 

Because the Hungarian algorithm has a $O(N^3)$  numerical cost, we tested this algorithm on systems with a small number of particles.
\begin{figure}[h]
\includegraphics[scale=0.3,angle=-90]{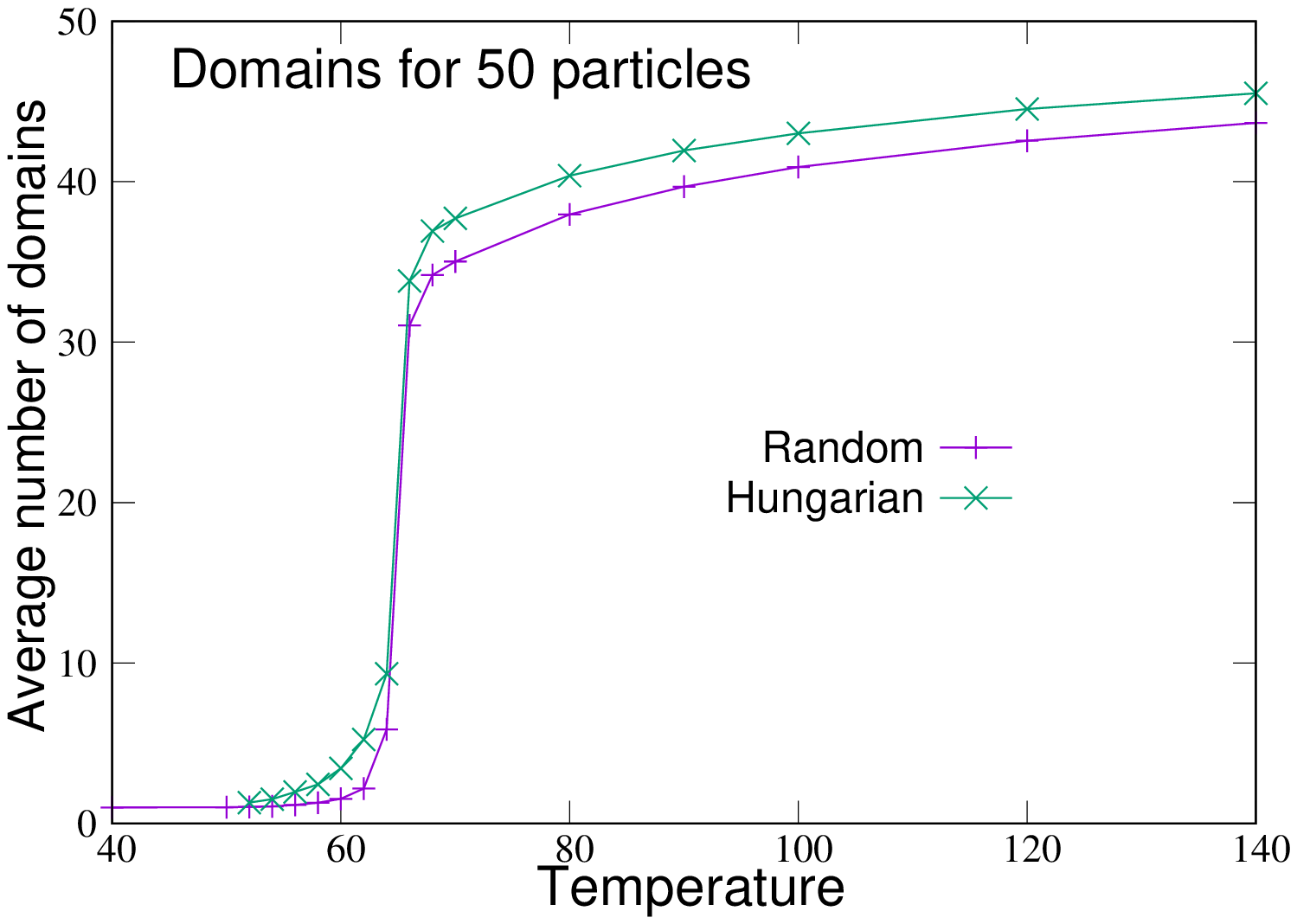}
\caption{Average number of domains for a box of 50 particles at different temperatures (K) using two pairing between the two replicas: a random (and constant) one and a one obtained with the Hungarian algorithm (recalculated at each step).}
\label{n50domdom}
\end{figure}
The Figure (\ref{n50domdom}) displays the effect of the Hungarian algorithm at different temperatures for a system with $50$ particles. We observe that the Hungarian algorithm increases systematically (at all temperatures) the number of domains. Interestingly this increase is however very small at all temperatures, and this leads us to use the inexpensive fixed random pairing for larger systems.
 
As we may expect the average number of domains is a increasing function of the temperature from low temperature where there is only one domain containing the full system to asymptotically $N$ domains (of one pair of particles) at high temperature. The two curves present the same shape with a brutal increase around the same temperature which will appear later to be the critical point, i.e. the divergence of the heat capacity. 
Therefore the evolution of the number of domains looks like to reflect the phase transition. 

\begin{figure}[htp]
  \begin{center}
\includegraphics[scale=0.34,angle=-90]{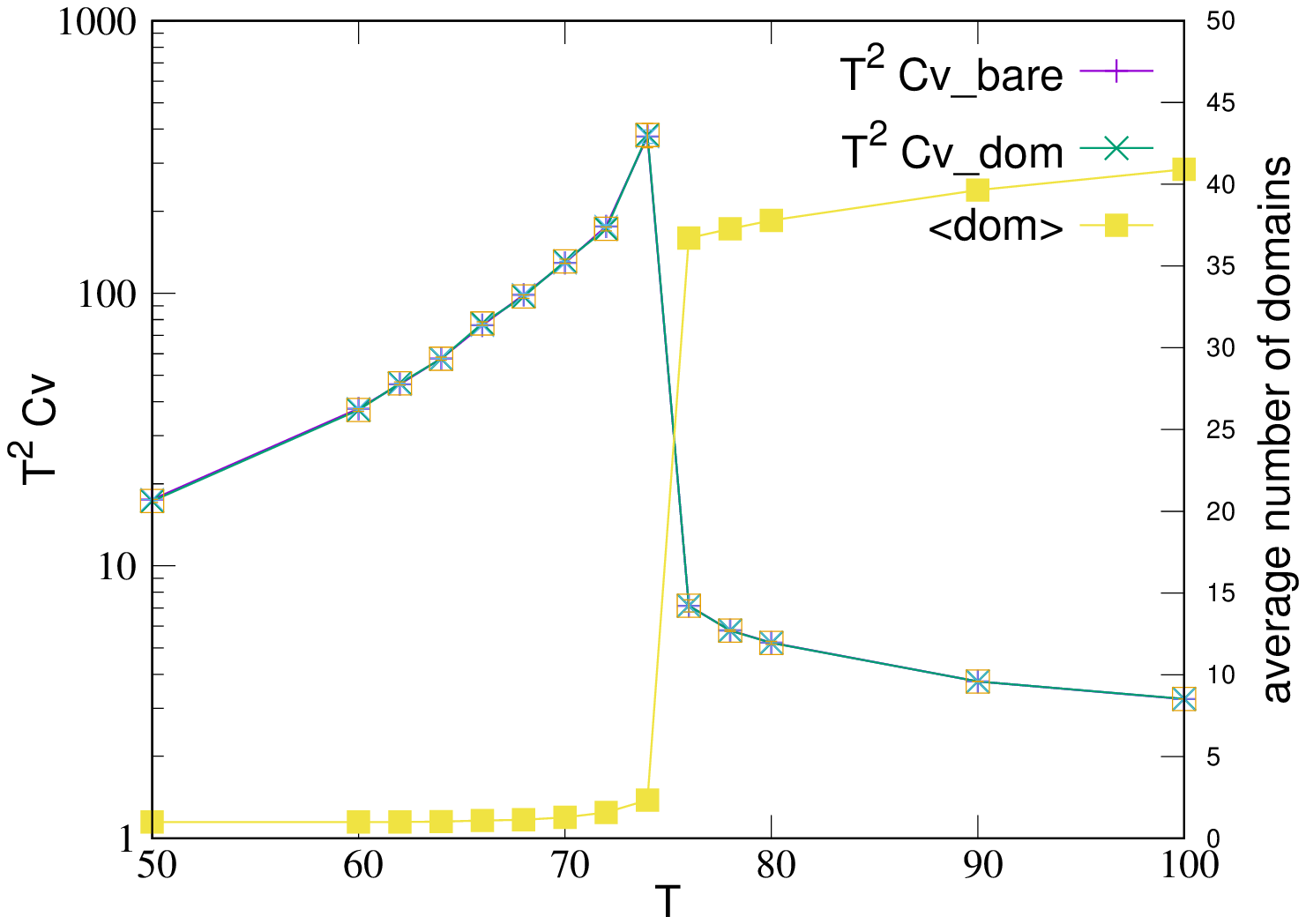}
\caption{Monte Carlo estimates of the expectation values of the two estimators variance of the Lennard Jones potential energy, i.e. $T^2Cv\_{\rm bare}$ and $T^2Cv\_{\rm dom}$ in function of the temperature T, and evolution of the average number of domains <dom>. The heat capacity estimates are provided using a logarithmic scale. }
\label{n50cvcvd_dom}
\end{center}
\end{figure}

Now we consider the two statistical averages~(\ref{stataverage}) involving $\chi$ and $\tilde{\chi}$ to compute the heat capacity 
\begin{eqnarray}
Cv\_{\rm bare}
& =& \frac{k}{T^2} (\overline{U^2} - \overline{U}^2)
\nonumber \\
Cv\_{\rm dom} & = & \frac{k}{T^2} \overline{\tilde{\chi}}
\end{eqnarray}
where in practice we use for the bare estimator $Cv\_{\rm bare}$ the two replicas, i.e. $U  = (\Uone+\Utwo)/2$.
The domain-exchange estimator $\tilde{\chi}$ is obtained with a fixed random pairing.
Figure (\ref{n50cvcvd_dom}) displays the estimates of the heat capacity  for Argon using a small box of 50 particles at various temperatures.

We checked out that the two estimators of the variance of $U$ are identical in the error bars for all temperatures which confirms numerically that the exchange-domain estimator is not biased. This unbiasedness is also apparent in the figure (\ref{n50cvcvd_dom}). 
In this figure, the curves of the heat-capacity and the average domain number display the same critical signature at a the same  temperature $T_c \simeq 60K$, which corresponds to a first order phase transition. 
We also observe a rapid and abrupt change around $Tc \sim 60 K$.
When $T<T_c$ only one domain (of size $N$) is selected,  when $T>T_c$ many domains of small sizes  are coming out. This suggests that the number of the domains gives an alternative accurate  criteria to detect a phase transition separating a finite from an infinite correlation length.
 The number of domains is the key parameter that should reduce the variance of $\tilde{\chi}$, thus the domain-exchange algorithm should be efficient when this number is large.

\begin{figure}
\begin{center}
\includegraphics[scale=0.32,angle=-90]{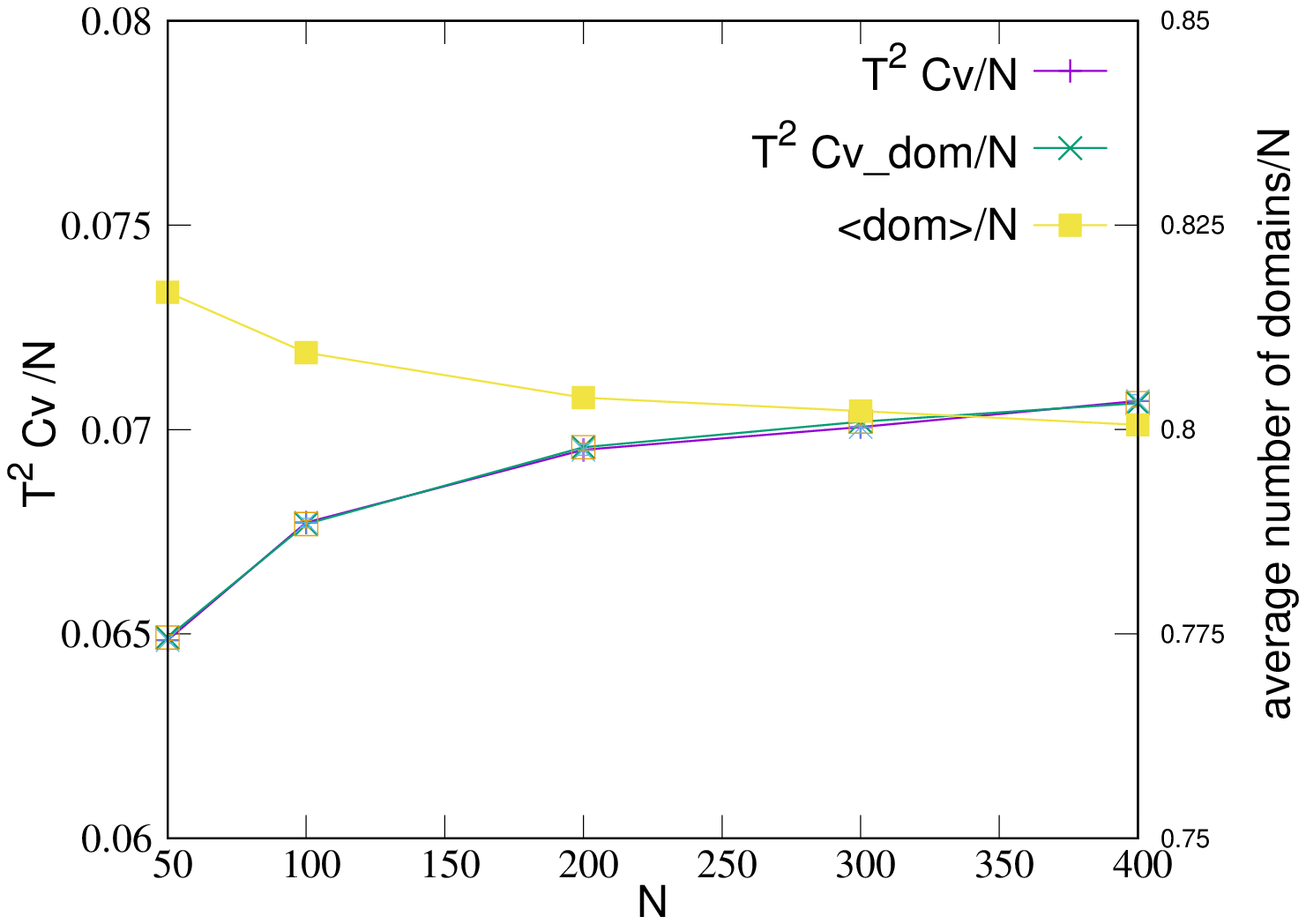}
\caption{Heat capacity per particle for the bare and exchange domain estimators at $T=100K$ as a function of the size $N$. Average number of domains per particle in the exchange domain algorithm at same temperature as a function of the size $N$.}
\label{relcv}
\end{center}
\end{figure}

\begin{figure}[!ht]
\begin{center}
\includegraphics[scale=0.32,angle=-90]{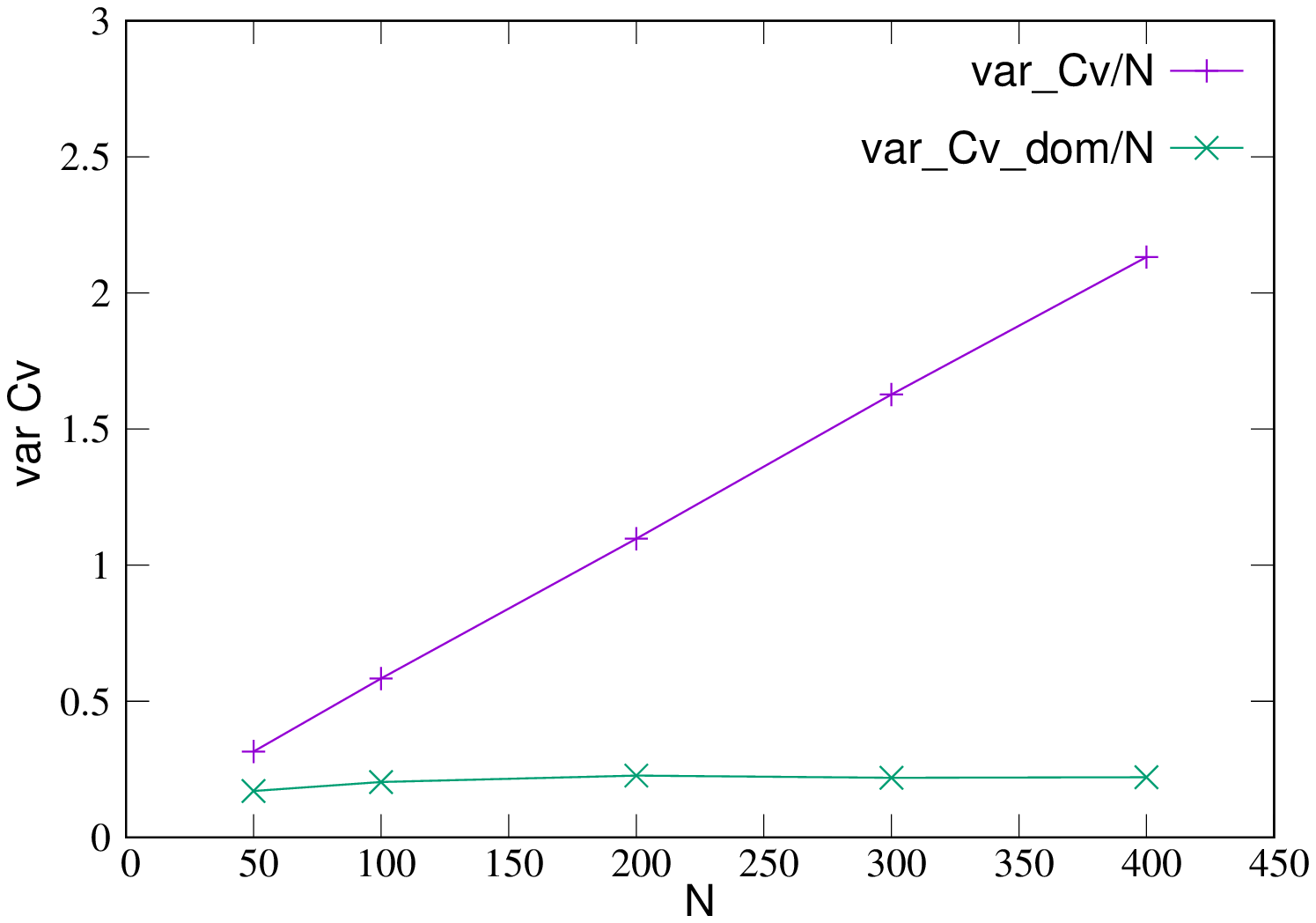}
\caption{ Variances per particle for the bare  and exchange domain estimators of $Cv$ as a function of the size $N$ at $T=100K$.}
\label{relcvvar}
\end{center}
\end{figure}
The graphs Fig.~(\ref{relcv}) and Fig.~(\ref{relcvvar}) display the average number of domains per particle, the heat capacity per particle and the variance per particle of its two estimators at $T=100K$.
The figure ~(\ref{relcv}) displays the saturation of the heat capacity per particle as a function of the size $N$ of the system, this, of course, illustrates the size-extensivity of $Cv$. This size-extensivity property applies clearly also to the number of domains, since this number per particle converges towards an asymptotic value.  
However the variance per particle of $Cv_{\rm bare}$ is increasing linearly with the system size, 
which illustrates the quadratic behavior of $Cv_{\rm bare}$.  
At the opposite the variance per particle of $Cv_{\rm dom}$ displays a very different behavior, it is nearly constant which illustrates the size-extensivity of the variance and  
the division by $N$ of the algorithm complexity in agreement with the theory.
 
\section{Conclusion and perspectives}
We introduced an algorithm to compute susceptibilities of a statistical system at finite temperature, to lower the scaling of the variance of the estimators by a factor $N$ near the thermodynamic limit. 
This has been obtained by introducing two independent replicas of the system, and sampling domains of particles which can be exchanged between the two replicas without modifying the Boltzmann distribution.
This method which extends the exchange domain algorithm can be used to write the susceptibility as an effective interaction energy between domains Eq.~(\ref{tchi}).
The advantage of this new estimator is that the size-extensivity of the susceptibility applies also to the variance of its new estimator. This property provides a gain of a factor $N$ close to the thermodynamic limit. 

We checked this property on a Lennard Jones model.
We observed two additional interesting results. The first is that the number of domains looks like to depend but very slightly on the pairing, i.e. the particles which have the same index, in the two replicas. 
This result is interesting numerically since we can avoid the costly Hungarian algorithm.
This is also interesting physically because clusters or domains may contain distant particles. In other words the clusters may be highly spatially delocalized. This is a different picture from usual cluster algorithms (applied on lattice spin models)  where the same sites of the two replicas are paired and the domains are spatially connected.

The second result is that the number of domains and the heat capacity displays the same  discontinuity signature at the same critical temperature.

This algorithm has been designed for 2-body interactions, however the extension to p-body interactions is in principle possible, for example  a  $4$-body interaction  $w(\R_i,\R_j,\R_k,\R_l)$ can be written as a two-body interaction between the variable $(\R_i,\R_j)$ and  $(\R_k,\R_l)$ or between  ${i}$ and ${j,k,l}$. 
Note that we have introduced a simple estimator with a size extensive variance, but the variance is still larger than the bare estimator when there is only one domain. This shows that there is room for improvement, the trivial one being a combination of both the bare and the new estimator. 
Finally generalization to high order logarithmic derivatives of the partition function, i.e  cumulants of higher than two order (covariances) representing the susceptibilities remains to be explored.

\appendix

\section{One-domain limit}
Let's compare the estimators of the covariance of two variables $U,V$ in the one-domain limit (in practice at small temperature when the only domain built by the algorithm is the full pair of replicas).

Supposing that $U$ and $V$ are centered, $U_1 V_1$ and $U_2 V_2$ are respectively unbiased estimators of the covariance on the first and second replicas.
A better estimator is the average 
\begin{equation}
X = \frac{1}{2} (U_1 V_1 + U_2 V_2)
\end{equation}
which we refer as the standard estimator. 
In the one-domain limit the exchange domain estimator is  
\begin{equation}
\tilde{X} = \frac{1}{2} (U_1 -U_2)(V_1-V_2) 
\end{equation}

We will prove that the variances of the two estimators are related as follows 
\begin{equation}
  \Var(\tilde{X}) = \Var(X) + \E(U^2) \E(V^2)
 \end{equation} 
In  other words the exchange domain estimator has a larger variance i.e. a lower  performance in the one domain limit.

Using the independence of $U_1 V_1$ and $U_2 V_2$ we find the variance of $X$
\begin{equation}
  \Var(X) = \frac{1}{2} V(UV) = \frac{1}{2}(\E(UV)^2-\chi^2)
\end{equation} 
 Considering that  
 $$\tilde{X} = X -\frac{1}{2}(U_1 V_2 + U_2 V_1)$$  

\begin{eqnarray}
  \Var(\tilde{X})-\Var(X) &=& \frac{1}{4}\Var(U_1 V_2 + U_2 V_1)-\cov(X,U_1 V_2+U_2 V_1) \nonumber 
 \end{eqnarray}
In this expression covariance can be found to be zero, this can be checked by developing its expression and using independence and centering properties.
 
 After developing the term which remains in the r.h.s. that is the variance we find  
 \begin{eqnarray}
  \Var(X) &=&   \frac{1}{2}(\Var(U)\Var(V)+\cov(U_1V_2, U_2 V_1)) \nonumber \\
      &=& \E(U^2) \E(V^2)
\end{eqnarray}
after using again independence and centering properties.
And of course 
\begin{equation}
 \Var(\tilde{X}) = \Var(X) + \Var(U) \Var(V)
\end{equation}    
if  $U$ and  $V$ are centered.



\bibliography{domain.bib}
\end{document}